\documentstyle[twocolumn,aps,epsf]{revtex}

\begin{document}
\draft
\preprint{HEP/123-qed}
\title{Phase Transitions in Quantum Pattern Recognition}
\author{C. A. Trugenberger}
\address{InfoCodex SA, av. Louis-Casai 18, CH-1209 Geneva, Switzerland\\
Theory Division, CERN, CH-1211 Geneva 23, Switzerland}
\address{e-mail: ca.trugenberger@InfoCodex.com}
\date{\today}
\maketitle
\begin{abstract}
With the help of quantum mechanics one can formulate a model of associative
memory with optimal storage capacity. I generalize this model by introducing a
parameter playing the role of an effective temperature. The corresponding
thermodynamics provides criteria to tune the efficiency of quantum pattern
recognition. I show that the associative memory undergoes a phase transition
from a disordered, high-temperature phase with no correlation between input and
output to an ordered, low-temperature phase with minimal input-output Hamming
distance. 
\end{abstract}

\narrowtext
The power of quantum computation \cite{review} is mostly associated with the
speed-up in computing time it can provide with respect to its classical
counterpart. Recently, however, I showed \cite{myselfa} that
this new paradigm of information processing opens the possibility for another
improvement upon classical computation, represented by associative memories with
exponential, and thus optimal, storage capacity. Subsequent further studies
\cite{scj} are turning quantum pattern recognition into a completely new
application of quantum information theory.

In traditional computers the storage
of information is address-oriented. Retrieval of
information requires a precise knowledge of the memory address and, therefore,
incomplete or noisy inputs are not permitted.
In order to address this shortcoming, models of associative (or
content-addressable) memories \cite{neuralnetworks} were introduced. Here,
recall of information is possible on the basis of partial knowledge of their
content, without knowing the storage location. The best known examples
are the Hopfield model and its generalizations \cite{hopfield}. 

While these models solve the problem of recalling incomplete or noisy inputs,
they suffer from a severe capacity shortage. Due to the phenomenon of crosstalk,
which is essentially a manifestation of the spin glass transition \cite{parisi}
in the corresponding spin system, 
the maximum number of binary patterns that
can be stored in a Hopfield network of $n$ neurons is linear in the
number of neurons, $p_{max} = O(n)$ \cite{neuralnetworks}.

The probabilistic associative quantum memory 
proposed in \cite{myselfa} solves both
problems. It is content-addressable and can thus recognize corrupted or
incomplete inputs and it can store $2^n$ binary patterns on $n$ qbits. Contrary
to its classical counterpart, which matches any input onto a stored pattern, the
quantum associative memory is characterized by both a recognition
process and an identification process. An input pattern can be rejected as
non-recognized even before an identification is attempted. For its simplest
version, described in \cite{myselfa}, the
identification efficiency cannot be tuned; only the recognition efficiency
can be influenced. 

In this paper I propose a generalization of my previous model with a new
parameter $t$ playing the role of an effective temperature, which can be tuned
by adding a number $b = [1/t]_{\rm integer}$ of certain control qbits. A proper
thermodynamics corresponding to this parameter $t$ can be defined. In
particular, the free energy $F(t)$ describes the average
behaviour of the recognition mechanism at temperature $t$ and provides
criteria to tune the efficiency of the associative memory. I show that, by
increasing $b$ (lowering $t$), the associative memory undergoes a phase
transition from a disordered phase with no correlation between input and output
to an ordered phase with minimal Hamming distance bewteen the input and the
output. This extends to quantum information theory the relation with Ising spin
systems known in error-correcting codes and in public key
cryptography \cite{addref}.

The memory model proposed in \cite{myselfa} consists of three registers: one for
the input, one for the memory $|m\rangle$ proper and one for a control qbit
$|c\rangle$. The memory $|m\rangle$ consists of a coherent superposition of the
$p$ binary patterns $|p^i\rangle$ on $n$ entangled qbits:
\begin{equation}
|m\rangle = {1\over \sqrt{p}} \ \sum_{k=1}^p |p^k\rangle \ .
\label{aa}
\end{equation}
I do not discuss here the algorithm for generating this superposition starting
from the $p$ individual patterns, since it is described in detail in
\cite{myselfa}. Also, the usual error-correcting infrastructure \cite{review} is
assumed in order to protect the memory from decoherence.

The information retrieval algorithm entails repeating a set of operations and
measurements of the control qbit $|c\rangle$ until this is found in state
$|0\rangle$ or a threshold $T$ of repetitions is reached. When $|c\rangle =
|0\rangle$ is measured one can proceed to a measurement of the memory register
that yields the output; if $T$ is reached before obtaining $|c\rangle
=|0\rangle$ the input is classified as ``non-recognized".

I propose here to generalize this device by increasing to $b$ the number of
control qbits and repeating sequentially all operations for each of them before
measuring the control register. The full initial quantum state is thus:
\begin{equation}
|\psi_0\rangle = {1\over \sqrt{p}} \sum_{k=1}^p |i; p^k;
0_1,\dots , 0_b\rangle
\label{ab}
\end{equation}
where $|i\rangle = |i_1,\dots ,i_n\rangle$ denotes 
the input qbits, the second register, $m$, contains the memory (\ref{aa})
and all $b$ control qbits are in
state $|0\rangle$. Applying the Hadamard gate \cite{review}
$H = \left( \sigma_1+\sigma_3 \right) /\sqrt{2}$ ($\sigma_i$ being the Pauli
matrices) to the first control qbit one obtains
\begin{eqnarray}
|\psi _1\rangle &&= {1\over \sqrt{2p}} \ \sum_{k=1}^p |i; 
p^k; 0_1,\dots ,0_b\rangle
\nonumber \\ 
&&+{1\over \sqrt{2p}} \ \sum_{k=1}^p |i; 
p^k; 1_1,\dots ,0_b\rangle \ .
\label{ad}
\end{eqnarray}
I now apply to this state the following combination of quantum gates:
\begin{equation}
|\psi _2\rangle = \prod_{j=1}^n \ NOT_{m_j} 
\ XOR_{i_j m_j} |\psi _1\rangle \ ,
\label{ae}
\end{equation}
where the single-qbit gate NOT is represented by the first Pauli matrix
$\sigma_1$, while the two-qbit exclusive OR (XOR) has the matrix representation
XOR = diag(1,$\sigma_1$) and performs thus a NOT on the second qbit if and only
if the first one is in state $|1\rangle$. Subscripts indicate the qbits on which
these gates are applied, $m$ denoting the memory register.

As a result of the above operation the memory register qbits
are in state $|1\rangle$ if $i_j$ and $p^k_j$ are identical
and $|0\rangle$ otherwise:
\begin{eqnarray}
|\psi _2\rangle &&= {1\over \sqrt{2p}} \ \sum_{k=1}^p |i; 
d^k; 0_1,\dots ,0_b\rangle 
\nonumber \\
&&+{1\over \sqrt{2p}} \ \sum_{k=1}^p |i; 
d^k; 1_1,\dots ,0_b\rangle \ ,
\label{af}
\end{eqnarray}
where $d^k_j = 1$ if and only if  $i_j=p^k_j$ and $d^k_j=0$ otherwise.

Consider now the following Hamiltonian:
\begin{eqnarray}
{\cal H} &&= \left( d_H \right)_m \otimes \left( \sigma_3 \right)_{c_1} \ ,
\nonumber \\
\left( d_H \right)_m && = \sum_{j=1}^n 
\left( {\sigma_3 + 1\over 2} \right) _{m_j}\ ,
\label{ag}
\end{eqnarray}
where $\sigma _3$ is the third Pauli matrix.
${\cal H}$ measures the number of 0's in register $m$, with a plus sign if $c_1$
is in state $|0\rangle$ and a minus sign if $c_1$ is in state $|1\rangle$. Given
how I have prepared the state $|\psi _2\rangle$, this is nothing else than the
number of qbits which are different in the input and memory registers $i$ and
$m$. This quantity is called the {\it Hamming distance} and represents the
(squared) Euclidean distance between two binary patterns. 

Every term in the superposition (\ref{af}) is an eigenstate of ${\cal H}$ with a
different eigenvalue. Applying thus the unitary operator ${\rm exp} 
(i \pi {\cal H}/2n)$ to $|\psi _2\rangle$ one obtains
\begin{eqnarray}
|\psi _3\rangle &&= {\rm e}^{i{\pi \over 2n}{\cal H}} \ |\psi_2\rangle \ ,
\label{ah} \\
|\psi_3\rangle &&= {1\over \sqrt{2p}} \sum_{k=1}^p {\rm e}^{i{\pi\over 2n}
d_H\left( i, p^k\right)}
|i; d^k; 0_1,\dots ,0_b\rangle 
\nonumber \\
&&+ {1\over \sqrt{2p}} \sum_{k=1}^p {\rm e}^{-i{\pi\over 2n}
d_H\left( i, p^k\right)}
|i; d^k; 1_1,\dots ,0_b\rangle \ ,
\nonumber
\end{eqnarray}
where $d_H\left( i, p^k \right)$ denotes the Hamming distance bewteen the input
$i$ and the stored pattern $p^k$. 

In the final step I restore the memory gate to the state 
$|m\rangle$ by applying
the inverse transformation to eq. (\ref{ae}) and I 
apply the Hadamard gate  
to the control qbit $c_1$, thereby obtaining
\begin{eqnarray}
|\psi _4\rangle &&= H_{c_1} \prod_{j=n}^1 XOR_{i_j m_j}  
\ NOT_{m_j} \ |\psi_3\rangle \ ,
\label{ai} \\
|\psi_4\rangle &&= {1\over \sqrt{p}} \sum_{k=1}^p {\rm cos}\  {\pi \over
2n} d_H\left( i, p^k\right) 
|i; p^k; 0_1,\dots ,0_b\rangle 
\nonumber \\
&&+ {1\over \sqrt{p}} \sum_{k=1}^p {\rm sin}\  {\pi \over 2n}
d_H\left( i, p^k\right) 
|i; p^k; 1_1,\dots ,0_b\rangle .
\nonumber
\end{eqnarray}

The idea is now to repeat the above operations sequentially for all $b$ control
qbits $c_1$ to $c_b$. This gives 
\begin{eqnarray}
|\psi_{\rm fin}\rangle &&= {1\over \sqrt{p}} \sum_{k=1}^p \sum_{l=0}^b
\ {\rm cos}^{b-l} \left( {\pi\over 2n} d_H\left( i, p^k \right)\right) \times 
\nonumber \\
&&{\rm sin}^l \left( {\pi\over 2n} d_H\left( i, p^k \right)\right) 
\ \sum_{\left\{ J^l \right\}} |i; p^k; J^l\rangle ,
\label{al}
\end{eqnarray}
where $\left\{ J^l \right\}$ denotes the set of all binary numbers of
$b$ bits with exactly $l$ bits 1 and $(b-l)$ bits 0.
This concludes the deterministic part of the information retrieval process.

At this point one needs a measurement 
of the control register. Note that the overall effect obtained by the
deterministic operations is an overall amplitude concentration on memory states
similar to the input if there is a large number of $|0\rangle$ control qbits and
an amplitude concentration on states different to the input if there is a large
number of $|1\rangle$ control qbits. One is thus interested in retaining the
projected state after the measurement only if all control qbits are measured in
state $|0\rangle$. This will generically entail repeating the deterministic part
of the algorithm several times, until exactly the desired state for the control
register is obtained. If the number of such repetitions exceeds a preset
threshold $T$ the input if classified as "non-recognized" and the algorithm is
stopped. Otherwise, once $|c_1, \dots , c_b\rangle = |0_1, \dots, 0_b\rangle$ is
obtained, one proceeds to a measurement of the memory register $m$, which yields
the output pattern of the memory. 

Since the expected number of repetitions needed to measure the desired control
register state is $1/P_b^{\rm rec}$, with
\begin{equation}
P_b^{\rm rec} = {1\over p} \ \sum_{k=1}^p \ {\rm cos}^{2b} \left(
{\pi \over 2n} d_H \left( i; p^k\right) \right) \ ,
\label{am}
\end{equation}
the probability of measuring $|c_1,\dots ,c_n\rangle = |0_1, \dots ,0_n\rangle$,
the threshold $T$ governs the {\it recognition efficiency} of the input
patterns.

Once the input pattern $i$ is recognized, the measurement of the memory register
yields the stored pattern $p^k$ with probability
\begin{eqnarray}
P_b\left( p^k\right) &&= {1\over Z} \ {\rm cos}^{2b} \left( {\pi \over 2n}
d_H\left( i, p^k\right) \right) \ ,
\label{an} \\
Z &&= pP_b^{\rm rec} = \sum_{k=1}^p {\rm cos}^{2b} \left( {\pi \over 2n}
d_H\left( i, p^k\right) \right) \ .
\label{ao}
\end{eqnarray}
Clearly, this probability is peaked around those patterns which have the
smallest Hamming distance to the input. The highest probability of retrieval is
thus realized for that pattern which is most similar to the input..

Contrary to the simplest version of this model presented in \cite{myselfa},
however, here there is a second tunable parameter, namely the number $b$ of
control qbits. This new parameter $b$ controls the {\it identification
efficiency} of the quantum memory since, increasing $b$, the probability
distribution $P_b\left( p^k\right)$ becomes more and more peaked on the 
low $d_H\left( i, p^k \right) $ states, until
$
\lim_{b\to \infty} P_b\left( p^k \right) = \delta_{k k_{\rm min}}\ ,
$
where $k_{\rm min}$ is the index of the pattern (assumed unique for convenience)
with the smallest Hamming distance to the input.

The role of the parameter $b$ becomes familiar upon a closer examination of 
eq.( \ref{an}). Indeed, the quantum distribution described by this equation is
equivalent to a canonical Boltzmann distribution 
with (dimensionless) temperature
$t = 1/b$ and (dimensionless) energy levels 
\begin{equation}
E^k = -2 \ {\rm log} \ {\rm cos} \left( {\pi \over 2n} d_H\left( i, p^k
\right) \right) \ ,
\label{aq}
\end{equation}
with $Z$ playing the role of the partition function. 

The appearance of an effective thermal distribution suggests studying the
average behaviour of quantum associative memories via the corresponding
thermodynamic potentials. Before this can be done, however, one must deal with
the different distributions of stored patterns characterizing each individual
memory. To this end I propose to average also over this distribution, by keeping
as a tunable parameter only the minimal Hamming distance $d$ between the input
and the stored patterns. In doing so, one obtains an average description of the
average memory.

As a first step it is useful to normalize the pattern representation 
by adding (modulo 2) to all
patterns, input included, the input pattern $i$. This clearly preserves all
Hamming distances and has the effect of normalizing the input to be the state
with all qbits in state $|0\rangle$. The Hamming distance $d_H\left( i, p^k
\right)$ becomes thus simply the number of qbits in pattern $p^k$ with value
$|1\rangle$. For loading factors $p/n \to 0$ in the limit $n\to \infty$
the partition function for the average memory takes a particularly simple form:
\begin{equation}
Z_{\rm av} = {p\over N_{\lambda}} 
\ \sum_{\{\lambda \}} \ \sum_{j=d}^n \ \lambda_j
\ {\rm cos}^{2b}\left( {\pi \over 2} {j\over n}\right) \ ,
\label{ar}
\end{equation}
where $\lambda_j$ describes an unconstrained probability 
distribution such that $\sum_{j=d}^n
\lambda_j = 1$, $\{ \lambda \}$ is the set of such distributions and
$N_{\lambda}$ the corresponding normalization factor.

I now introduce the free energy $F(b,d)$ by the usual definition
\begin{equation}
Z_{\rm av} = p \ {\rm e}^{-bF(b,d)} = Z_{\rm av}(b=0) \ {\rm e}^{-bF(b,d)} \ ,
\label{as}
\end{equation}
where I have chosen a normalization such that ${\rm exp}(-bF)$ describes the
deviation of the partition function from its value for $b=0$ (high effective
temperature). Since $Z/p$, and consequently also $Z_{\rm av}/p$ 
posses a finite, non-vanishing large-$n$ limit,
this normalization ensures that $F(b,d)$ is intensive, exactly like the energy
levels (\ref{aq}), and scales as a constant for large $n$. This is the only
difference with respect to the familiar situation in statistical mechanics.

The free energy describes the equilibrium of the system at effective temperature
$t=1/b$ and has the usual expression in terms of the internal energy $U$ and the
entropy $S$:
\begin{eqnarray}
F(t,d) &&= U(t,d) - tS(t,d) \ ,
\nonumber \\
U(t,d) && = \langle E \rangle _t \ ,\quad 
S(t,d) = {-\partial F(t,d) \over \partial t} \ .
\label{at}
\end{eqnarray}
Note that, with the normalization I have chosen in (\ref{as}), 
the entropy $S$ is
always a negative quantity describing the deviation from its maximal value 
$S_{\rm max} = 0$ at $t=\infty$.

By inverting eq.(\ref{aq}) with $F$ substituting $E$ one can also define an
effective (relative) input/output Hamming distance ${\cal D}$ 
at temperature $t$:
\begin{equation}
{\cal D}(t,d) = {2\over \pi} \ {\rm arccos} \ {\rm e}^{-F(t,d)\over 2} \ .
\label{au}
\end{equation}
This corresponds exactly to representing the recognition probability 
of the average memory as
\begin{equation}
\left( P_b^{\rm rec} \right) _{\rm av} = 
{\rm cos}^{2b} \left( {\pi \over 2} {\cal D}(b,d) \right) \ ,
\label{av}
\end{equation}
which can also be taken as the primary definition of the effective Hamming
distance. 

The function ${\cal D}(b,d)$ provides a complete description of the
behaviour of quantum associative memories, which can be used to tune their
performance. Indeed, suppose that one wants the memory to recognize and identify
inputs with up to $\epsilon n$ corrupted inputs with an efficiency of $\nu$
$(0\le \nu \le 1)$. Then one must choose a number $b$ of control qbits
sufficiently large that 
$\left( {\cal D}(b,\epsilon n) - \epsilon \right) \le 
\left( 1-\nu \right)$ and a
threshold $T$ of repetitions satisfying $T \ge 1/{\rm cos}^{2b} \left(
{\pi \over 2} {\cal D}(b,\epsilon n) \right) $, as illustrated in Fig. 1 below.

A first hint about the general behaviour of the effective distance function
${\cal D}(b,d)$ can be obtained by examining closer the energy eigenvalues
(\ref{aq}). For small Hamming distance to the input these reduce to
\begin{equation}
E^k \simeq {\pi^2\over 4} \left( {d_H\left( i, p^k\right) \over n}
\right) ^2\ ,\qquad {d_H\left( i, p^k \right) \over n} \ll 1\ .
\label{aw}
\end{equation}
Choosing again the normalization in which $|i\rangle = |0 \dots 0\rangle$ and
introducing a ``spin" $s_i^k$ with value $s_i^k = -1/2$ if qbit
$i$ in pattern $p^k$ has value $|0\rangle$ and $s_i^k=+1/2$ if qbit $i$ in
pattern $p^k$ has value $|1\rangle$, one can express the energy levels for 
$d_H/n \ll 1$ as
\begin{equation}
E^k = {\pi ^2\over 16} + {\pi ^2\over 4n^2}
\sum_{i,j} s_i^k s_j^k + {\pi ^2 \over 4n} \sum_i s_i^k \ .
\label{ay}
\end{equation}
Apart from a constant, this is the Hamiltonian of an infinite-range
antiferromagnetic Ising model in presence of a magnetic field. The
antiferromagnetic term favours configurations $k$ with half the spins up 
and half down, so that $s^k_{\rm tot} =\sum_i s^k_i = 0$, giving $E^k =
\pi^2/16$.The magnetic field, however, tends to align the
spins so that $s^k{\rm tot} = -n/2$, giving $E^k = 0$. Since this is
lower than $\pi^2/16$, the ground state configuration is ferromagnetic, with all
qbits having value $|0\rangle$. At very low temperature (high $b$), where the
energy term dominates the free energy, one expects thus an ordered phase of the
quantum associative memory with ${\cal D}(t,d) = d/n$. This corresponds to a
perfect identification of the presented input. As the temperature is raised
($b$ decreased) however, the thermal energy embodied by the entropy term in the
free energy begins to counteract the magnetic field. At very high temperatures
(low $b$) the entropy approaches its maximal value $S(t=\infty) = 0$ (with the
normalization chosen here). If this value is approached faster than $1/t$, the
free energy will again be dominated by the internal energy . In this case,
however, this is not any more determined by the ground state but rather equally
distributed on all possible states, giving
\begin{eqnarray}
F(t=\infty) &&= U(t=\infty) = {-1\over 1-{d\over n}} 
\int_{d\over n}^1 \ dx
\ 2\ {\rm log} \ {\rm cos}\left( {\pi \over 2} x\right) 
\nonumber \\
&&= \left( 1+{d\over n}
\right) 2 \ {\rm log }2 + O\left( \left( {d\over n}\right) ^2\right) \ ,
\label{az}
\end{eqnarray}
and leading to an effective distance
\begin{equation}
{\cal D}(t=\infty, d) = {2\over 3} - {2\ {\rm log }2 \over \pi \sqrt{3}} 
\ {d\over n} + O\left( \left( {d\over n}\right) ^2\right) \ .
\label{azz}
\end{equation}
This value corresponds to a disordered phase with no correlation between input
and output of the memory. 

A numerical study of the thermodynamic potentials in (\ref{at}) and (\ref{au})
indeed confirms a phase transition from the ordered to the disordered phase as
the effective temperature is raised. In Fig. 1 I show the effective
distance ${\cal D}$ and the entropy $S$ 
for 1 Mb ($n=8 \times 10^6$) patterns and
$d/n= 1\%$ as a function of the inverse temperature $b$ (the entropy is rescaled
to the interval [0,1] for ease of presentation). At high temperature there is
indeed a disordered phase with $S=S_{\rm max} =0$ and ${\cal D} = 2/3$. At
low temperatures, instead, one is in the ordered phase with $S=S_{\rm min}$ and
${\cal D}=d/n=0.01$. The effective Hamming distance plays thus the role of the
order parameter for the phase transition. 

\begin{figure}
\centerline{\epsfysize=8.5cm\epsfbox{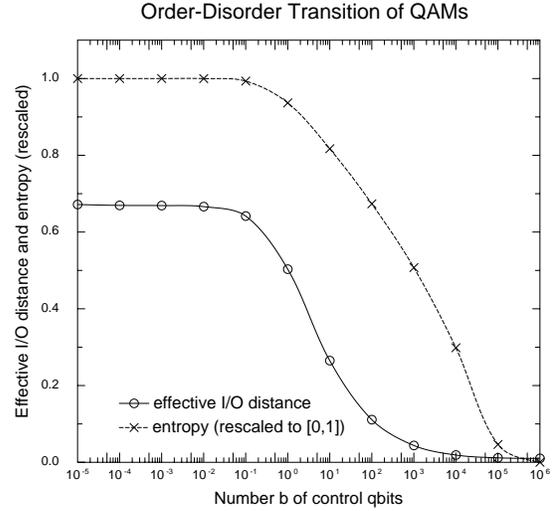}}
\caption{Effective input/output distance and entropy (rescaled to [0,1]) 
for 1Mb patterns and $d/n = 1\%$.}
\end{figure}

The phase transition occurs at $b_{\rm cr} \simeq 10^{-1}$. The physical regime
of the quantum associative memory ($b$ = positive integer) begins thus just
above this transition. For a good accuracy of pattern recognition one should
choose a temperature low enough to be well into the ordered phase. As is clear
from Fig. 1, this can be achieved already with a number of control qbits
$b=O(10^4)$. Note that this number becomes independent of the dimension $n$
of the patterns for large $n$. The computational load of quantum pattern
recognition is thus determined uniquely by the accuracy requirements.

\end{document}